\documentclass[a4paper,final]{paper}  % A nice looking "paper"
\usepackage[utf8]{inputenc}
\usepackage[UKenglish]{babel}

\usepackage{amsmath}
\makeatletter
\renewcommand*\env@matrix[1][*\c@MaxMatrixCols c]{%
  \hskip -\arraycolsep
  \let\@ifnextchar\new@ifnextchar
  \array{#1}}
\makeatother

\usepackage{amsfonts}
\usepackage{amssymb}

\usepackage{graphicx}
\usepackage[small]{caption} % Extend the formats for captions
\usepackage{subcaption}

\usepackage{array}
\usepackage{multirow}

\usepackage{url} % To add url-format to the document
\usepackage{mathtools} % More tools than those offered by "amsmath"
\usepackage{color} % Package to work with colours
\usepackage[usenames,dvipsnames,svgnames,table]{xcolor}

\usepackage[ruled]{algorithm} % Support for algorithms
\usepackage{algpseudocode} % Required to use "algorithmic" environment
\usepackage{csquotes} % Add this, otherwise a warning is thrown if not
\usepackage{listings}

\newcommand{\vect}[1]{\mathbf{#1}}

\usepackage{authblk}

\title{Software System Design based on Patterns for Newton-Type Methods}

\author[ ,a]{Ricardo Serrato Barrera\thanks{rsbserrato@inaoep.mx}}
\author[ ,b]{Gustavo Rodr\'{i}guez
  G\'{o}mez\thanks{grodrig@inaoep.mx}}
\author[ ,b,c]{Julio C\'{e}sar P\'{e}rez
  Sansalvador\thanks{jcp.sansalvador@inaoep.mx, correspondence author}}
\author[ ,b]{Saul E. Pomares Hern\'{a}ndez\thanks{spomares@inaoep.mx}}
\author[ ,d]{Leticia Flores Pulido\thanks{leticia.florespo@udlap.mx}}
\author[ ,e]{Antonio Mu\~noz\thanks{jose.munoz@cucsur.udg.mx}}
\affil[a]{Estratei Sistemas de Informaci\'{o}n S.A. de C.V., Virrey de Mendoza 605-B, Col. Las fuentes, 59699, Zamora, Michoac\'{a}n, M\'{e}xico}
\affil[b]{Instituto Nacional de Astrof\'{i}sica, \'{O}ptica y
  Electr\'{o}nica, Computer Science Department, Luis Enrique Erro 1, 72840, Tonantzintla, Puebla, M\'{e}xico}
\affil[c]{C\'atedra CONACyT - Instituto Nacional de Astrof\'{i}sica, \'{O}ptica y
  Electr\'{o}nica, Computer Science Department, Luis Enrique Erro 1, 72840, Tonantzintla, Puebla, M\'{e}xico}
\affil[d]{Universidad Aut\'{o}noma de Tlaxcala, Facultad de Ciencias
  B\'{a}sicas, Ingenier\'{i}a y Tecnolog\'{i}a, Calzada Apizaquito,
  Colonia Apizaquito, 90300, Apizaco, Tlaxcala, M\'{e}xico}
\affil[e]{Departamento de Ingenier\'ias, Universidad de Guadalajara,
  Ave. Independencia Nacional, Autl\'an, Jalisco, M\'exico}

\date{ }                    % No date to appear

\providecommand{\keywords}[1]{\textbf{\textit{Keywords---}} #1}

\begin{document}

\maketitle
%\onecolumn \maketitle \normalsize \vfill

% \begin{abstract}
%   The popularity of Newton's method due to its fast convergence rate
%   has led to the development of software tools implementing variations
%   of Newton's method. These tools are mainly used by the engineering,
%   mathematical and scientific community for the solution of nonlinear
%   and optimisation problems. The software implementations of
%   Newtons--type methods are mainly developed following the procedural
%   programming paradigm which tends to deliver code that is hard to
%   modify or extend. In order to address these issues we developed a
%   pattern-object-oriented software design that captures the structure
%   of Newton's type methods. We present the methodology that guided us
%   through the development of these software design. This methodology
%   has been previously applied for the development of scientific
%   software during the identification and application of design
%   patterns that fit the software design requirements. The novel
%   software design resulting from the application of the presented
%   methodology performs as a framework to facilitate the development
%   and implementation of new Newton--type methods.
% \end{abstract}

\begin{abstract}
  A wide range of engineering applications uses optimisation
  techniques as part of their solution process. The researcher uses
  specialized software that implements well-known optimisation
  techniques to solve his problem. However, when it comes to develop
  original optimisation techniques that fit a particular problem the
  researcher has no option but to implement his own new method from
  scratch. This leads to large development times and error prone code
  that, in general, will not be reused for any other application. In
  this work, we present a novel methodology that simplifies, fasten
  and improves the development process of scientific software. This
  methodology guide us on the identification of design patterns. The
  application of this methodology generates reusable, flexible and
  high quality scientific software. Furthermore, the produced software
  becomes a documented tool to transfer the knowledge on the
  development process of scientific software. We apply this
  methodology for the design of an optimisation framework implementing
  Newton's type methods which can be used as a fast prototyping tool
  of new optimisation techniques based on Newton's type methods. The
  abstraction, re-useability and flexibility of the developed
  framework is measured by means of Martin's metric. The results
  indicate that the developed software is highly reusable.

  % We develop a pattern-object-oriented software design that captures
  % the structure of Newton's type methods. We present the design
  % process and the methodology that guides us through the development
  % process of a software framework that facilitates the implementation
  % of new Newton--type methods. The main focus is in the flexibility of
  % the software design to integrate new Newton related
  % methods. Implementation strategies such as multicore or parallel
  % technologies to improve the performance of the developed software
  % are in current investigation.
\end{abstract}


%\begin{multicols}{2}

\section{Introduction}
\label{sec:introduction}
% Nota de acuerdo al libro de Higham usar e.g. es un estilo malo

A wide range of applications are solved using a variation of Newton's
method \cite{Barrera:2017:IP}, these include: trust regions methods
and line search methods \cite{Nocedal:1999:BK}, damping methods,
inexact or truncated methods \cite{Nash:2000:AR}, quasi--Newton
methods \cite{Xu:2001:AR}, or hybrid methods \cite{Blue:1980:AR},
\cite{Powell:1970:AR}. Each variation presents specific features that
make it suitable for the solution of a particular problem:
unconstrained optimisation, solution of nonlinear systems of
equations, or data fitting problems.

%\idea{Discussing software implementing Newton--type methods}

\begin{table}[h]
  \footnotesize
  %\centering
   \begin{tabular}{p{2.5cm} c p{6.8cm}}
  %  \begin{tabular}{ccp{6.8cm}}
     \hline\noalign{\smallskip}
     Name & Language & Comments\\
     \noalign{\smallskip}\hline\noalign{\smallskip}
     MINPACK 1.0 \cite{More:1980:TR} & Fortran & Optimization package\\
     MINOS 5.0 \cite{Murtagh:1983:TR} & Fortran & Optimization package\\
     UNCMIN \cite{Schnabel:1985:AR} & Fortran & Unconstrained
                                                minimisation, evolved
                                                to TENSOLVE\\
     HOMPACK \cite{Watson:1987:AR} & Fortran & Homotopies package\\
     TENSOLVE \cite{Bouaricha:1997:AR} & Fortran & Tensor methods package\\
     NITSOL \cite{Pernice:1998:AR} & Fortran & Nonlinear equations solver package\\
     Matlab software routines \cite{Kelley:2003:BK} & Matlab & Nonlinear equations solvers\\
     COOOL \cite{Deng:1996:BK} & \texttt{C++} & Optimization and matrix operations package\\
     PETSc \cite{Balay:2018:RT} & \texttt{C++} & Differential equations and
                                                 nonlinear systems solver
                                                 package\\
     OPT++ \cite{Meza:2007:AR} & \texttt{C++} & Optimization package\\
     \noalign{\smallskip}\hline
   \end{tabular}
   \caption{A selection of software packages implementing
     Newton--type methods to solve specialized problems.}
   \label{Table:software_newtons}
\end{table}

Currently, there is a large collection of specialized software
packages that implement variations of Newton--type methods; see Table
\ref{Table:software_newtons}. Note that most of these software
packages are implemented following the \textit{procedural programming
  paradigm} which main drawback is the lack of flexibility of the
developed code; that is, the software is hard to extend or reuse
\cite{Meza:2007:AR,Bruaset:1997:AR,
  Dongarra:1994:IP,Gockenbach:1999:AR,Matthey:2004:AR}. Generally,
software packages developed using the procedural programming paradigm
present the following problems:

\begin{itemize}
\item Code that is difficult to read and understand due to the use of
  unsuitable data structures, \cite{Dongarra:1994:IP}.
\item Code that does not describe the algorithms that implements,
  \cite{Gockenbach:1999:AR}.
\item Code that is hard to reuse or modify since it is based on very
  restrictive programming concepts and the large number of parameters
  in the interfaces \cite{Bruaset:1997:AR,Matthey:2004:AR}.
\end{itemize}

%\idea{Netwon type methods oo software}

Only three of the software packages presented in Table
\ref{Table:software_newtons} implement \textit{the object-oriented
  programming} approach; this shows us the preference of the
procedural programming paradigm over the object-oriented paradigm in
the scientific community. The three software packages implementing the
object-oriented approach are: \texttt{COOOL} (The CWP -Center for Wave
Phenomena- Object-Oriented Optimization Library) \cite{Deng:1996:BK},
\texttt{PETSc} (Portable, Extensible Toolkit for Scientific
Computation) \cite{Balay:2018:RT} and \texttt{OPT++}
\cite{Meza:2007:AR}. \texttt{COOOL} only handles unconstrained
optimisation problems and do not fully exploit the object-oriented
paradigm; concepts such as inheritance and polymorphism are not
exploited in its software design. \texttt{PETSc} implements Newton's
method for the solution of nonlinear systems of equations, it is a set
of routines and data structures for the solution of PDEs that is
widely used for the scientific community. \texttt{OPT++} is a software
library for nonlinear optimisation in \texttt{C++}, it uses
inheritance and polymorphism to provide an interface hierarchy to
implement quasi--Newton methods, inexact Newton methods, line search
methods, and trust region methods.

%\idea{Introducing object--oriented approach}

The application of object-oriented techniques for the development of
scientific software started in the late 90's; \texttt{Diffpack}
\cite{Bruaset:1997:AR} was a pioneer project in the application of
object-oriented techniques, it was a software for the simulation of
engineering and scientific applications.

%\idea{Introducing the benefits gained by the application of design
% patterns}

% In the words of Gamma \textit{et al.} \cite{Gamma:1995:BK} ``The
% design of this type of software is mainly a creative process; the
% more experience of the software architect the more reusable the
% software solution is''.

\subsection{Software Design Patterns and Scientific Software
  Development}
\label{subsec:software_desing_patterns}
Software design patterns were firstly introduced by Gamma \textit{et
  al.} in the book ``\textit{Design patterns: elements of reusable
  object--oriented software}'' \cite{Gamma:1995:BK}. Design patterns
are software design solutions to common and repeatable software design
problems. Software design patterns benefit low coupling and strong
cohesion relations between software components; therefore, the
resulting software is easy to reuse, modify and extend.

% Enterprise software developers apply software design patterns as part
% of their development process since design patterns capture the
% knowledge of experienced software developers for the solution of
% design problems. In recent years, the scientific community has shown
% interest in the application of design patterns to benefit from the
% knowledge of expert software designers for the development of
% scientific software
% \cite{Dongarra:1994:IP,Gockenbach:1999:AR,Matthey:2004:AR,Meza:2007:AR,Rouson:2010:AR,Rouson:2011:BK,Filippone:2012:AR,Barbieri:2012:IP}.

% In the review literature, there is no mention of a process or
% methodology that explains how to determine the software patterns
% that better solve the design problem. In this work we propose a
% methodology and because there is not a manual or methodology it is
% here where most of the design problems arise} \DEB{this is how we
% link with the next paragraph

% Here, the problem is to identifying the patterns that able to solve
% a scientific problem since there is not methodology that aids to
% choose them and the related work does not explain their process for
% considering their patterns.

% In this section, we present a collection of articles whose authors
% develop software for scientific applications.  Some works propose
% scientific computing patterns and other use enterprise software
% patterns to develop their scientific applications.

The application of design patterns for the development of business
software dates back to the 1990's \cite{Rouson:2011:BK}. Even though,
currently there is neither a recognised pattern language for the
numerical specialists nor a research area focused on the specification
or application of design patterns for the development of scientific
software.

In recent years, the scientific community has shown interest in the
application of design patterns to benefit from the knowledge of expert
software designers for the development of scientific software
\cite{Meza:2007:AR,Dongarra:1994:IP,Gockenbach:1999:AR,Matthey:2004:AR,Rouson:2011:BK,
  Rouson:2010:AR,Filippone:2012:AR,Barbieri:2012:IP}. Tables
\ref{Table:propose_new_scientific_patterns} and
\ref{Table:use_patterns_in_scientific_computing} show a list of
selected works in scientific computing that apply design patterns, we
organised them into two main categories:
\begin{itemize}
\item Works that \textit{find, create and specify} new scientific
  software patterns, see Table
  \ref{Table:propose_new_scientific_patterns}.\footnote{According to
    Vlissides \cite{Vlissides:1998:BK}, any new proposed scientific
    pattern can not be immediately recognised as real patterns until
    they are validated by specialist, and their offered solutions are
    applied in other contexts.}
\item Works that \textit{borrow, modify and apply} software patterns
  from other areas (such as those for the development of
  administrative software) for the development of scientific software,
  see Table
  \ref{Table:use_patterns_in_scientific_computing}
\end{itemize}

\begin{table}
  \footnotesize
  % \centering
  \begin{tabular}{p{2.0cm} p{3.4cm} p{5.2cm}}
    \hline\noalign{\smallskip}
    Work & Context & Proposed patterns\\
    \noalign{\smallskip}\hline\noalign{\smallskip}
    Blilie, 2002, \cite{Blilie:2002:AR} & Dynamic systems & Particle--Particle, Particle--Mesh, Mesh.\\
    Rodriguez et al. 2004, \cite{Rodriguez:2004:IP} & Dynamic
                                                      systems & Model-Solver, System-Modularization\\
    Matthey et al. 2005, \cite{Cickovski:2005:TR} & Nodes
                                                    three-dimensional
                                                      simulation of
                                                    morphogenesis,
                                                    Molecular
                                                    Dynamics & Generic
                                                               Automation,
                                                               Plugins,
                                                               Dynamic
                                                               Class
                                                               Nodes,
                                                               Policy,
                                                               Multigrid,
                                                               Strategy
                                                               Chain\\
    Heng and Mackie 2009, \cite{Heng:2009:AR} & Finite Element &
                                                                 Model-Analysis
                                                                 separation,
                                                                 Model-UI
                                                                 separation,
                                                                 Modular
                                                                 element,
                                                                 Composite
                                                                 element,
                                                                 Modular analyzer\\
    Rouson et al. 2010, \cite{Rouson:2010:AR} & Multiphysics
                                                Modeling & Semi-Discrete, Surrogate, Puppeteer\\
    Rouson et al. 2011, \cite{Rouson:2011:BK} & Multiphysics
                                                Modeling & Abstract
                                                           Calculus,
                                                           Surrogate,
                                                           Puppeteer\\
    \hline\noalign{\smallskip}
  \end{tabular}
  \caption{Works that propose new scientific software patterns.}
  \label{Table:propose_new_scientific_patterns}
\end{table}

% Their work exemplify how to abstract common operations of
% mathematical entities, and how to change integration strategies.
\begin{table}
  \footnotesize
  \centering
  \begin{tabular}{p{5.0cm} p{5.0cm}}
    \hline\noalign{\smallskip}
    Work & Context\\
    \noalign{\smallskip}\hline\noalign{\smallskip}
    Padula et al. 2004, \cite{Padula:2004:TR} & Simulation and
                                                optimization\\
    Decyk et al. 2008, \cite{Decyk:2008:AR} & Particle in cell
                                              simulation\\
    Sansalvador et al. 2011, \cite{PerezS:2011:IP} & Multirate
                                                     integration
                                                     methods\\
    Rouson et al. 2011, \cite{Rouson:2011:BK} & Multiphysics
                                                Modeling\\
    Barbieri et al. 2012, \cite{Barbieri:2012:IP} & Matrix
                                                    Computations\\
    Filippone et al. 2012, \cite{Filippone:2012:AR} & Matrix
                                                      Computations\\
    Barrera et al. 2017, \cite{Barrera:2017:IP} & Newton-type
                                                  methods software\\
    \hline\noalign{\smallskip}
  \end{tabular}
  \caption{Works that uses software patterns in scientific software
    applications.}
  \label{Table:use_patterns_in_scientific_computing}
\end{table}

The works listed in Table
\ref{Table:use_patterns_in_scientific_computing} benefit from the
application of design patterns in their software designs, they gain
flexibility and reusability. Nevertheless, none of them explains the
methodology used to select the software patterns.

\subsubsection{Difficulties on the application of design patterns for
  the development of scientific software}
The specification of a pattern is often abstract, hence determining or
mapping a pattern to a specific application is a difficult task that
mainly relies on the expertise of the software developer. In order to
apply a design pattern, we must \textit{identify the software elements
  that comprise the design}.

In the scientific software field there are no explicit relations
between numerical concepts and design patterns, therefore we need to
establish such relations in order to develop a numerical software
based on software patterns. In the particular case of Newton--type
methods, we have to identify or establish relations between the
abstract concepts in the domain of the problem, the numerical methods
involved in the implementation of Newton--type methods, and the design
pattern's domain.

The main contribution of this work is a methodology that guides the
researcher through the identification of software design patterns to
efficiently apply the object-oriented paradigm to develop reusable and
easy to adapt software. This methodology applies the object--oriented
paradigm, domain analysis, and Scope Commonality and Variability
analysis (SCV) \cite{Coplien:1998:AR}. We start by performing a domain
analysis that help us to generate abstractions to understand the
problem domain in terms of software patterns. Then we select a subset
of design patterns from Gamma's book \cite{Gamma:1995:BK} that can be
applied to address the problems in the software design. We provide
guidelines to determine what software patterns to use. Finally, we
apply Martin's metrics \cite{Martin:2003:BK} to provide evidence that
the resulting software design is easy to reuse, extend and modify.

We apply the mentioned methodology to develop a framework for the
implementation of Newton--type methods. We capture the structure of
Newton--type methods in a novel software design by means of software
patterns. Our newly developed software system design facilitates the
implementation of algorithms by providing a flexible and extensible
framework to incorporate new Newton--type methods and easily integrate
off-the-shelf software libraries.

\subsection{Paper Organisation}
\label{subsec:paper_organisation}
In Section~\ref{sec:mathematical_problem} we introduce the
mathematical notation used in the paper and define the mathematical
problem regarding Newton--type methods. Then in
Section~\ref{sec:methodology} we describe the methodology used to
identify concepts and relations to generate software components that
will be associated via design patterns. The application of this
methodology for the development of a Newton--type methods software
system is presented in Section~\ref{sec:application}. In
Section~\ref{sec:newton_type_methods_software_evaluation} we present
an evaluation of the resulting software design by using Martin's
metrics. The conclusions and future work are presented in
Section~\ref{sec:conclusions}.

% The purpose of this section is to present the mathematical problem
% to be considered
\section{The mathematical problem}
\label{sec:mathematical_problem}
We start by identifying the main concepts in problems involving
Newton--type methods. We use the following notation throughout this
work. The $n$-dimensional Euclidean space is denoted by $\mathbb{R}^n$. A
vector $\vect{x} \in \mathbb{R}^n$ is to be understood as a column
vector. The vector $\vect{x}_* \in \mathbb{R}^n$ denotes a solution, and
$\{\vect{x}_k\}, k = 0,1,\ldots, M$ is a sequence of iterates. The $i$-th
component of the vector $\vect{x}_k$ is denoted by
$\vect{x}_{k,i}$. The gradient of the function $f(\vect{x})$ is
denoted by
$\nabla f(\vect{x}) = (\partial f/ \partial
\vect{x}_1,\ldots, \partial f / \partial \vect{x}_n)^T$. The Hessian
of $f(\vect{x})$ is the matrix
$Hf(\vect{x}) = (\partial^2 f/\partial \vect{x}_i \partial
\vect{x}_j)_{i,j}$ and the Jacobian of the function $F(\vect{x})$ is
the matrix $JF(\vect{x}) = (\partial F_i/\partial \vect{x}_j)_{i,j}$,
where $F_i(\vect{x})$ are the $i$-th component functions of
$F(\vect{x})$. We use the Euclidean norm:
$\lVert \vect{x} \rVert^2 = \sum_{i=1}^n \vect{x}_i^2$.

We consider three classes of nonlinear problems that appear in many
applications of the real--world.
\begin{itemize}
\item The nonlinear equations problem or \textbf{NE}, which comprises
  to find ${\vect{x}}_*$ such that the vector-value function $F$ of
  $n$ variables satisfies $F({\vect{x}}_*)={0}$.
\item The unconstrained optimisation problem or \textbf{UO}, which
  involves to find ${\vect{x}}_*$ such that the real-value function
  $f$ of $n$ variables satisfies $f({\vect{x}}_*) \leq f({\vect{x}})$
  for all ${\vect{x}}$ \textit{close to} ${\vect{x}}_*$.
\item The nonlinear least-square problem or \textbf{NLS}, which
  requires to find ${\vect{x}}_*$ for which
  $\sum_{i=1}^{m}(r_i(\vect{x}))^2$ is minimised, where
  $r_i(\vect{x})$ denotes the $i$-th component function of
  $R(\vect{x}) = (r_1(\vect{x}), r_2(\vect{x}), \ldots,
  r_m(\vect{x}))^T$, $\vect{x}\in \mathbb{R}^n$, $m \ge n$.
\end{itemize}

The above problems are are mathematically equivalent under reasonable
hypotheses \cite{Dennis:1996:BK}. For example, the \textbf{NE} problem
can be transformed into the \textbf{UO} problem by using the Euclidean
norm and defining the $f: \mathbb{R}^n \rightarrow \mathbb{R}$ function as
\begin{equation}
  f = \frac{1}{2} \lVert F(\vect{x}) \rVert^2 
\end{equation}

For the above problems a Newton--type method is involved, in Algorithm
\ref{Alg:generic_newton} we present the generic form of a Newton
algorithm given by Kelley in \cite{Kelley:1999:BK}.
\begin{algorithm}
  \caption{Generic Newton Algorithm}
  \label{Alg:generic_newton}
  \begin{algorithmic}[1]
    \While{Stopping condition is not satisfied}
    \State $s \gets$ calculate Newton direction
    \State $\lambda \gets$ calculate step length
    \State $\vect{x}_{k+1}=\vect{x}_{k}+\lambda s$
    \EndWhile
  \end{algorithmic}
\end{algorithm}
where $s$ is called the Newton direction and $\lambda$ the step
length. Different methods exist to compute these variables, for
example, Newton's method, Quasi-Newton's method, Newton's method with
Cholesky decomposition, method of Steepest Descent, Line Search
methods, Trust Region methods, among others.

\section{Methodology for patterns identification}
\label{sec:methodology}
One of the key stages when using design patterns for software
development is the identification of relations between software
components. In the case of scientific software development there are
no explicit relations between software patterns and numerical
methods. The methodology presented in this section helps us to
identify and establish relations between the abstract concepts in the
problem domain and the software patterns.

The main tasks of our proposed methodology are the following:
\begin{itemize}
\item Find key concepts of the problem domain.
\item Perform an SCV analysis to identify software components that
  remain invariant through different scenarios, and to identify
  software components that may change at run time.
\item Analyse the resulting software components and their relations to
  identify software design problems and recognize software design
  patterns that may solve these problems.
\end{itemize}

The above listed tasks may be decomposed in the following steps:
\begin{enumerate}
\item Create a \textit{general description of the problem} to identify
  key concepts of the problem domain. These concepts define the
  \textit{Scope} of the SCV analysis.
\item Study the \textit{commonalities and variabilities} of the
  concepts in different scenarios by applying the \textit{Analysis
    Matrix approach} by Shalloway; see
  \cite{Shalloway:2004:BK}. Variations of the concepts generally lead
  to different versions or implementations of software
  components. These variations are generally encapsulated to
  facilitate change and adaptation of the developed software.
\item Find the \textit{most important concepts or participants}, and
  study their relations. These concepts may represent subsystems in
  the final software design.
\item \textit{Identify complex relations} between concepts that may
  lead to software design problems.
\item Recognize \textit{software desing patterns} thay may be applied
  to solve the design problems found in the previous step.
\end{enumerate}

\begin{figure}[h]
  \centering
  \includegraphics[width=1.0\textwidth]{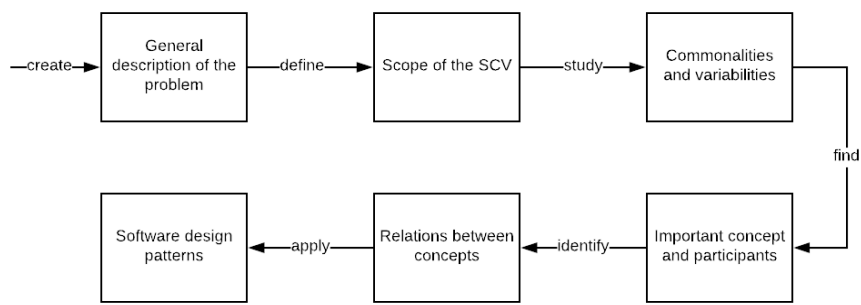}
  \caption{Software patterns identification methodology.}
  \label{Fig:software_patterns_identification_methodolgy}
\end{figure}

In Figure \ref{Fig:software_patterns_identification_methodolgy} we
highlight the key concepts of the above presented methodology.

%%%%%%%%%%Observaciuon GRG %%%%%%%%%%%%%%%%%%%%%
%%% Podría ser conveniente que en la figura 1 en la segunda caja incluyas Scope y SCV.

% Variations in the concepts are generally encapsulated to facilitate

% The identification of This helps us to identify relations between
% concepts at an early stage of the design process.

% Variations are generally encapsulated to facilitate change and
% adaptation of the developed software. We can use encapsulation to deal
% with variants, however, we Different versions of the software can be
% componThose components that variateVariantions Once variations have
% been identified in the problem we can think of encapsulation as a way
% to deal with those variants.

% Present the architectural design in function of the domain problem.

\section{Software design for Newton--type methods}
\label{sec:application}
\subsection{UML and the class diagrams}
\label{sec:uml_and_the_class_diagramas}
We use class diagrams from the Unified Modelling Language (UML) to
present our developed software design. UML diagrams are widely used in
the software development community. There are different type of
diagrams, each providing specific type of information. In our case, we
are interested in the relations between the objects that compose our
software design, therefore we use class diagrams which help us in
visualising these relations.

A class is represented by a rectangle with the class name in the upper
half of the rectangle. The lower half is used to specify methods and
properties associated with the class. \footnote{The methods and
  properties of the classes in our software design are not shown to
  focus the attention in the relations between the classes.}

There are two main type of relationships between classes:
\begin{itemize}
\item An \textbf{is-a} relation, When one class is a \textit{sub-type}
  of another class.
\item A \textbf{has-a} relation, when one class \textit{uses} or
  \textit{contains} another class.
\end{itemize}

\begin{figure}[h]
  \centering
  \includegraphics[width=0.5\textwidth]{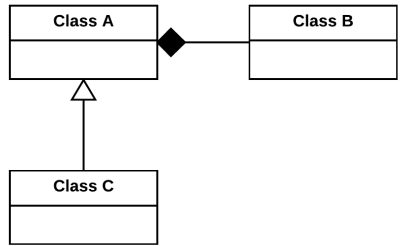}
  \caption{A simple class diagram showing the two main relationships
    between classes.}
  \label{Fig:class_diagram_simple}
\end{figure}

In Figure \ref{Fig:class_diagram_simple} we present three classes,
each represented by a rectangle. The line with the white triangle
connecting \texttt{Class A} and \texttt{Class C} represents an
\textit{is-a} relation. The \texttt{Class C} is a sub-type or
sub-class of \texttt{Class A}. The line with the black diamond
connecting \texttt{Class A} with \texttt{Class B} represents an
\textit{has-a} relation. In this case we have a \textit{composition}
relation, it means that when \texttt{Class A} is deleted then
\texttt{Class B} is deleted as well. A white diamond represents an
\textit{aggregation} relation, if that would be the case then
\texttt{Class A} would not be responsible for the life-cycle of
\texttt{Class B}.

\subsection{Software system design foundations}
\label{subsec:software_system_foundations}
At this stage, our main goal is to generate the foundations of the
software design. These foundations should allow us to add and modify
software components easily. A software design that supports easy
incorporation and exchange of Newton--type methods will help
researches when comparing the performance of these type of methods.

We start by analysing the generic form of a Newton--type method
presented in Algorithm \ref{Alg:generic_newton} and performing a
top-down analysis of the iterative form of the method. Newton--type
methods share the following basic sequence:
\begin{equation}
  x_{k+1}=x_{k}+\lambda s,
\end{equation}
where $x_{k+1}$ is equal to $x_{k}$ \textit{shifted} by $\lambda
s$. The \textbf{step length} $\lambda$ is calculated to guarantee
convergence of the algorithm, and the \textbf{Newton direction} $s$ is
found with a linear or quadratic model based on Taylor series. The
step length and the Newton direction are components of Newton--type
methods. These components will represent subsystems in the whole
system design \cite{Sommerville:2011:BK}. The variations of these
components in different scenarios are summarised in Table
\ref{Table:SCV_for_newtons}.

%$\lambda_* = \argmin_{\lambda>0} f(x+\lambda s)$
%%%%%%% GRG %%%%%%
%%% En la siguiente tabla no está especificado el concepto para Line search method
%%%%%%%%%%%%%
\begin{table}[h]
  \footnotesize
  % \centering
  \begin{tabular}{p{2.4cm} p{4.0cm} p{4.0cm}}
    \hline\noalign{\smallskip}
    \textbf{Scenarios} & \textbf{Concept}: Newton direction $s$ & \textbf{Concept}: Step length $\lambda$\\
    \noalign{\smallskip}\hline\noalign{\smallskip}
    Line search methods & Solve $\varphi(\lambda) = f(x+\lambda s)$ with fixed $x$ and $s$ & Calculate an step length to guarantee convergence.\\
    Trust region methods & Solve a linear or quadratic model within
                           a trust ratio to find the Newton
                           direction. & The step length is implicit
                                        in the Newton direction.\\
    Damped methods & Solve a system of equations using a direct
                     method to find the Newton direction. & The step
                                                            length
                                                            is
                                                            found using Line search methods.\\
    Quasi-Newton methods & Solve the system $Ax=b$, where $A$ is an
                           approximation of the first or second
                           derivative. & Use Line search of Trust region
                                         methods to compute the step length.\\
    Inexact or truncated methods & Use an iterative method to solve
                                   the system $Ax=b$. & Use a Line
                                                        search
                                                        method to
                                                        calculate the step length.\\
    Hybrid methods & Find a trajectory between a Newton direction
                     and a gradient direction. & Use a Trust region
                                                 method to compute the step length.\\
    \hline\noalign{\smallskip}
  \end{tabular}
  \caption{Scope-Commonality and Variability (SCV) Analysis for Newton
    methods.}
  \label{Table:SCV_for_newtons}
\end{table}

The variation in Newton components produces different Newton methods,
for example, a \textit{damped Newton method} is instantiated by
varying the Step length and the Newton direction components as
presented in Table \ref{Table:SCV_for_newtons}.

We identified two additional components to those detailed above, the
\textbf{Stopping condition}, and the \textbf{evaluation
  function}. Even though these two additional Newton components do not
define Newton's method classes they are part of the iterative method
and must be considered in the software design.

\begin{figure}
  \centering
  \includegraphics[width=0.80\textwidth]{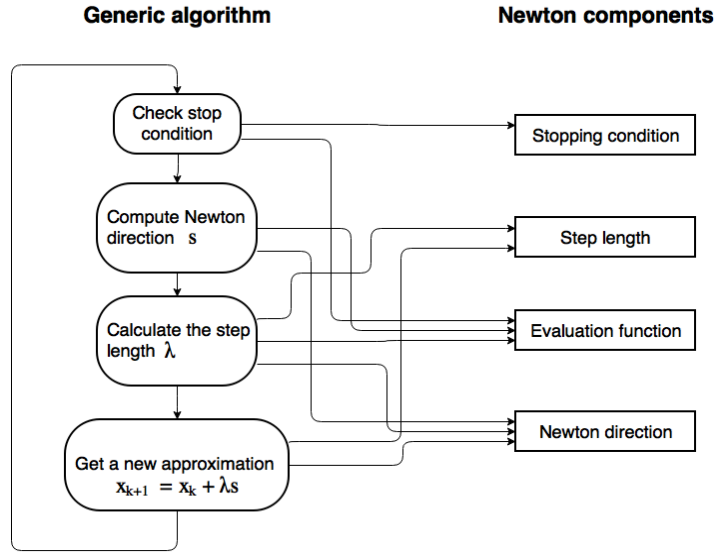}
  \caption{Relations between Newton components and the generic form of
    the Newton--type method.}
  \label{Fig:relations_concepts_algorithm}
\end{figure}

In Figure \ref{Fig:relations_concepts_algorithm} we present the
relations between the identified concepts and the steps of the generic
form of the Newton--type method. Each component represents a
subsystem. The step length component represents a subsystem
specialized in the searching step lengths.

Once the problem has been stated and the SCV analysis has been
performed, we proceed to identify problematic relationships and
interfaces in the software design. Consider the relations presented in
Figure \ref{Fig:relations_concepts_algorithm} and the generic form of
a Newton algorithm. Note that we can construct an specialized version
of a Newton-type method by specifying the components of each step of
the generic algorithm. In order to capture this structure, we apply
the \textbf{Template method} pattern \cite{Gamma:1995:BK} which
defines the skeleton of an algorithm and allows for variations at each
of its steps. We apply the \textbf{Facade} design pattern
\cite{Gamma:1995:BK} to provide a simple and general interface for the
Newtons components. This facilitates the variation of Newtons
components in the skeleton of the algorithm.

In order to support different implementations\footnote{Consider the
  case that the software developer provides a debugging and release
  implementation that share a common interface} of a single Newton
component we apply the \textbf{Bridge} design pattern
\cite{Gamma:1995:BK}, this pattern decouples the abstraction from the
implementation and lets them vary independently.

\begin{figure}[h]
  \centering
  \includegraphics[width=1.0\textwidth]{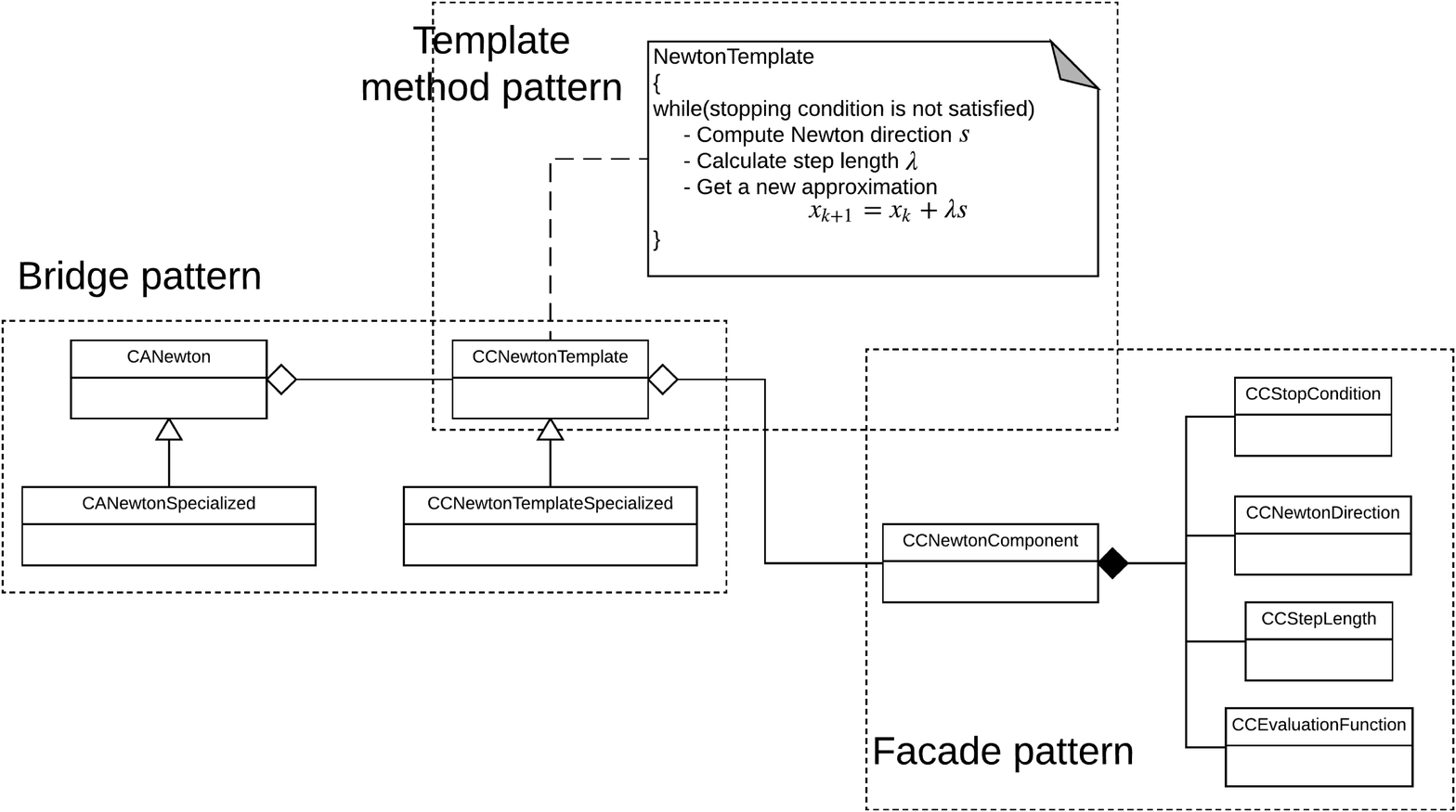}
  \caption{Foundations of the software design. We highlighted the
    application of the design patterns Template Method, Facade and
    Bridge in the software design. In the classes names the prefixes
    \texttt{CA} and \texttt{CC} indicate \textit{abstract} and
    \textit{concrete} classes, respectively.}
  \label{Fig:bridge_template_facade}
\end{figure}

The foundation components of the software design are described in
Figure \ref{Fig:bridge_template_facade}.

%See Figure \ref{Fig:bridge_template_facade} where we present the
%foundation components of the software design.

In the following sections we present the software design for some of
the identified Newton components.

\subsection{Software Design for Newton components}
\label{subsec:software_design_for_newton_components}
\subsubsection{Nonlinear Problems and the State pattern}
\label{subsubsec:nonlinear_problems_and_the_state_pattern}
\begin{figure}
  \centering
  \includegraphics[width=0.4\textwidth]{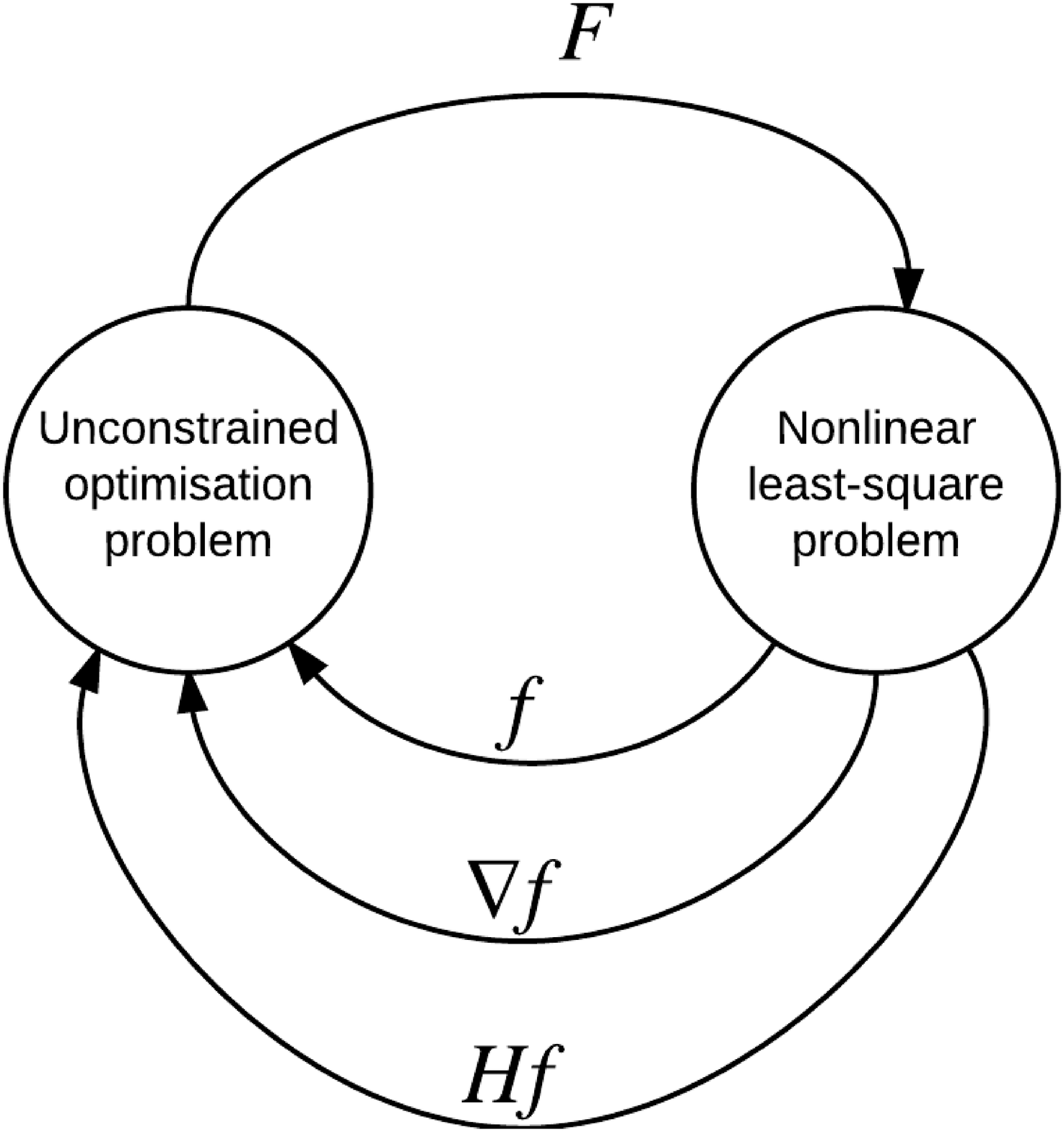}  
  \caption{Transition from an unconstrained optimisation problem to an
    nonlinear least-square problem, and \textit{vice versa}.}
  \label{Fig:state_newton_methods}
\end{figure}

The numerical schemes for solving the three nonlinear problems
presented in Section~\ref{sec:mathematical_problem} are closely
related. The nonlinear equation problem and the nonlinear least-square
problem are a particular case of an unconstrained minimization
problem. Given a vector-valued function ${F(\vect{x}) = 0}$ of $n$
variables we define $f(\vect{x}) = (1/2 )\rVert F(\vect{x})
\lVert^2$. Finding $\vect{x}_*$ such that $F(\vect{x}_*) = 0$ is
equivalent to find $\vect{x}_*$ such that $f(\vect{x}_*) = 0$. In
other words, a nonlinear least-square problem may be
\textit{transformed} into an unconstrained optimisation problem. This
transformation is represented as a state machine depicted in Figure
\ref{Fig:state_newton_methods}.

% \begin{figure}
% \centering
%    \includegraphics[scale=0.38]{old_figures/TransitionsUMNLS2}  
%    \caption{From an unconstrained optimisation problem to an nonlinear
%      least-square problem, and \textit{vice versa}.}
%   \label{Fig:Transitions_UMNLS2}
% \end{figure}

In each \textit{state} the nonlinear functions and its derivatives are
handled accordingly with the nonlinear problem represented by the
state.

Using the Analysis Matrix approach by Shalloway
\cite{Shalloway:2004:BK} and the SCV analysis we identified the
concepts and their variations in different scenarios as indicated in
Table \ref{Table:SCV_for_nonlinear_problems}.

\begin{table}[h]
  \footnotesize
  % \centering
  \begin{tabular}{p{1.4 cm} p{2.4 cm} p{2.4 cm} p{4.0 cm}}
    \hline\noalign{\smallskip}
    \textbf{Scenarios} & \textbf{Concept}: Nonlinear equations problem &
                                                                         \textbf{Concept}:
                                                                         Unconstrained
                                                                         optimisation problem
    & \textbf{Concept}: Nonlinear least-square problem\\
    \noalign{\smallskip}\hline\noalign{\smallskip}
    Function & $F:\mathbb{R}^{n}\rightarrow\mathbb{R}^{n}$ &
                                                             $f:\mathbb{R}^{n}\rightarrow\mathbb{R}$ & $F:\mathbb{R}^{n}\rightarrow\mathbb{R}^{m},n<m,$\\
                       & & & $f=\frac{1}{2}\lVert F \rVert ^{2}$\\
    First derivative & $JF(x)$ & $\nabla f(x)$ & $\nabla f(x)=$ $JF(x)^{T}F$\\
    Second derivative & N/A & $Hf(x)$ & $Hf(x)=JF(x)^{T} JF(x) +
                                         \sum_{i=1}^m F_{i} (H F_{i}(x))$\\
    \hline\noalign{\smallskip}
  \end{tabular}
  \caption{SCV analysis for nonlinear problems.}
  \label{Table:SCV_for_nonlinear_problems}
\end{table}

\begin{figure}
  \centering
  \includegraphics[width=0.8\textwidth]{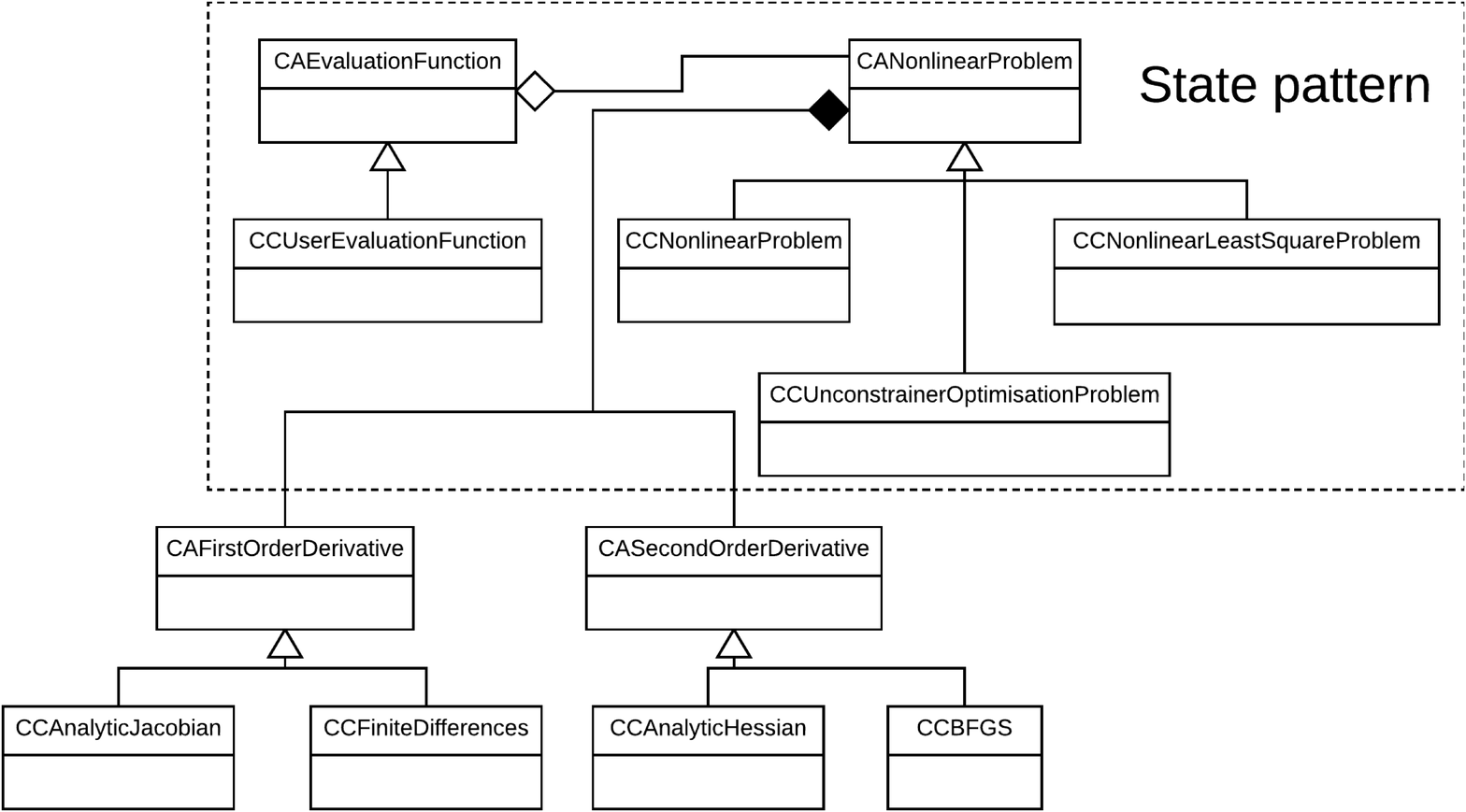}
  \caption{Software system design implementing transitions between
    nonlinear problems by means of the State pattern.}
  \label{Fig:state_pattern}
\end{figure}

Note that for a particular problem we compute either the Gradient, or
the Jacobian matrix of the function $F(x)$, or the Hessian matrix of
the function $f(x)$, \textit{i.e.} the derivative of the function
changes according to the nonlinear problem to solve. Changing the
nonlinear problem involves different operations to the functions
$f(x)$ or $F(x)$. The \textbf{State} design pattern
\cite{Gamma:1995:BK} implements an state machine to changes the
behaviour of an specific object. In our case, changing the nonlinear
problem produce a transition in the state machine that switches
between strategies to handle the nonlinear function and its
derivatives; see Figure \ref{Fig:state_pattern}.

% The procedure interfaces that contain the algorithms to calculate the
% derivatives of the function must be easy to deal with\DEB{modify,
%   adapt, update, understand?????} and to extend, such that we can
% introduce different algorithms to calculate the derivatives of the
% function. For example, we can implement their analytic form, compute
% them by finite differences or by quasi-Newton methods updates. Based
% on the analysis presented in Table
% \ref{Table:analysis_matrix_nonlinear_problems}, we define the criteria
% to define derivative interfaces and distinguish between two family
% classes, one for first order derivatives and another one for second
% order derivatives. These families of classes are used by the State
% pattern to switch between different methods to calculate derivatives
% based on the addressed nonlinear problem, see Figure
% \ref{Fig:state_pattern}.

% \begin{figure}
%   \centering
%    \includegraphics[scale=0.38]{old_figures/statepattern} 
%   \caption{State Pattern.}
%   \label{Fig:state_pattern}
% \end{figure}

% \DEB{Did not get the need of using the state pattern or what design
%   problem it is solving.}

\subsubsection{Line Search Methods and the Strategy pattern}
\label{subsubsec:line_search_methods_and_the_strategy_pattern}
The aim of line search methods is to find a search direction $s_k$ and
its corresponding step length $\lambda_k$. Typical strategies in line
search methods test a sequence of candidate values for $s$ and stop
when one of them satisfies some condition, for example when
$f(x_k + \lambda_k s_k) < f(x_k)$. Popular strategies are Wolfe,
Curvature and Goldstein conditions
\cite{Nocedal:2006:BK,Dennis:1996:BK}. These strategies involve two
main tasks: the computation of the step length, and the step length
decrease condition. In order to perform the SCV analysis we regard
these two tasks as concepts. The variations of these concepts are
presented in Table \ref{Table:SCV_line_search_methods}.

\begin{table}[h]
  \footnotesize
  % \centering
  \begin{tabular}{p{2.6 cm} p{3.8 cm} p{3.8 cm}}
    \hline\noalign{\smallskip}
    \textbf{Scenario} & \textbf{Concept}: Step length computation & \textbf{Concept}: Step length decreasing condition\\
    \noalign{\smallskip}\hline\noalign{\smallskip}
    \multirow{3}{*}{Line Search Methods} & Bisection & Wolfe\\
     & Quadratic interpolation & Goldstein\\
     & Cubic interpolation & Curvature condition\\
    \hline\noalign{\smallskip}
  \end{tabular}
  \caption{Variations of the concepts in the Line Search Methods
    scenario.}
  \label{Table:SCV_line_search_methods}
\end{table}

\begin{figure}[h]
  \centering
  \includegraphics[width=0.8\textwidth]{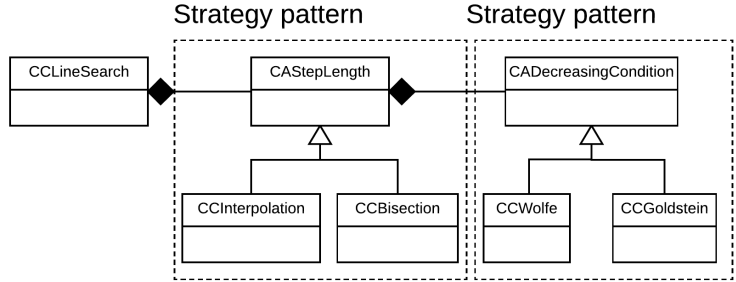}
  \caption{Software design for Line Search Methods applying a double
    Strategy Pattern for the implementation of the test condition and
    the step length computation procedure.}
  \label{Fig:double_strategy_pattern}
\end{figure}

The step length test condition is a sub-step in the computation of the
step length. The test condition can vary independently of the
selection of the strategy to calculate the step length. In order to
allow variations in the test condition separately from the step length
computation procedure we apply the \textbf{Strategy} pattern
\cite{Gamma:1995:BK}. This pattern defines a family of algorithms,
encapsulate them and makes them interchangeable. We apply a
\textit{double strategy}, one to encapsulate the step length test
condition, and the second one to encapsulate the method for the
computation of the step length, see Figure
\ref{Fig:double_strategy_pattern}.

The software design presented in this section belongs to the Newton
component: Step length from Figure
\ref{Fig:relations_concepts_algorithm}.

\subsubsection{Trust Region Methods and the Adapter pattern}
\label{subsub:trust_region_methods_and_the_adapter_pattern}
Trust region methods construct a mathematical model to approximate the
function $f(\vect{x})$ in a region around $\vect{x}_k$. The size of
this region, known as the \textit{trust regions}, is critical for the
effectiveness of each step. It is necessary to find a balance between
an small and a big region, therefore these methods look for a maximum
trust region expansion such that the model provides a good
approximation to the function $f(\vect{x})$. We have a constrained
minimization problem as follows:

\begin{equation}
  \label{eq:constrained_min_problem}
  \underset{s \in \mathbb{R}^n} \min \,m(s) = f({\vect{x}}_{k})+
  \nabla f({\vect{x}}_{k})^{T} s + \frac{1}{2} s^T Hf(\vect{x}_{k}) s
\end{equation}

\noindent subject to $\lVert s \rVert \leq \Delta$, where $\Delta > 0$ is the
trust region radius; see Section~\ref{sec:mathematical_problem} for
details on the mathematical notation.

The main tasks in trust region methods are the solution of the system
of equations derived from \eqref{eq:constrained_min_problem}, the
trust region radius update, and the model--function approximation. In
Table \ref{Table:SCV_analysis_trust_region_methods} we present the
variations of each of the tasks or concepts identified above.

\begin{table}[h]
  \footnotesize
  %\centering
  \begin{tabular}{p{1.6cm} p{2.8 cm} p{2.8 cm} p{2.8 cm}}
    \hline\noalign{\smallskip}
    \textbf{Scenario} & \textbf{Concept}: Solution of system of equations &
                                                                            \textbf{Concept}:
                                                                            Update
                                                                            trust
                                                                            region
                                                                            radio
    & \textbf{Concept}: Model--function approximation\\
    \noalign{\smallskip}\hline\noalign{\smallskip}
    Trust region & Cauchy point method & & \\
    methods & \textit{Dogleg} methods & Adaptive methods using thresholds \cite{Dennis:1996:BK}
     & Decreasing condition in Line Search Methods\\
     & Two--dimensional sub-space minimisation
                        methods & & \\
    \hline\noalign{\smallskip}
  \end{tabular}
  \caption{SCV analysis for Trust Region Methods.}
  \label{Table:SCV_analysis_trust_region_methods}
\end{table}

The identified concepts are highly related: a) if the trust region
radius is decreased too much then the descending direction found to
solve equation \eqref{eq:constrained_min_problem} may not be correct,
b) if the solution of the equation \eqref{eq:constrained_min_problem}
is not good enough then there might be many updates of the trust
region to increase or decrease the radius, and c) the trust region
update depends on how good is the approximation to the function
$f(\vect{x})$ given by the model (model--function approximation).

\begin{figure}
  \centering
  \includegraphics[width=0.8\textwidth]{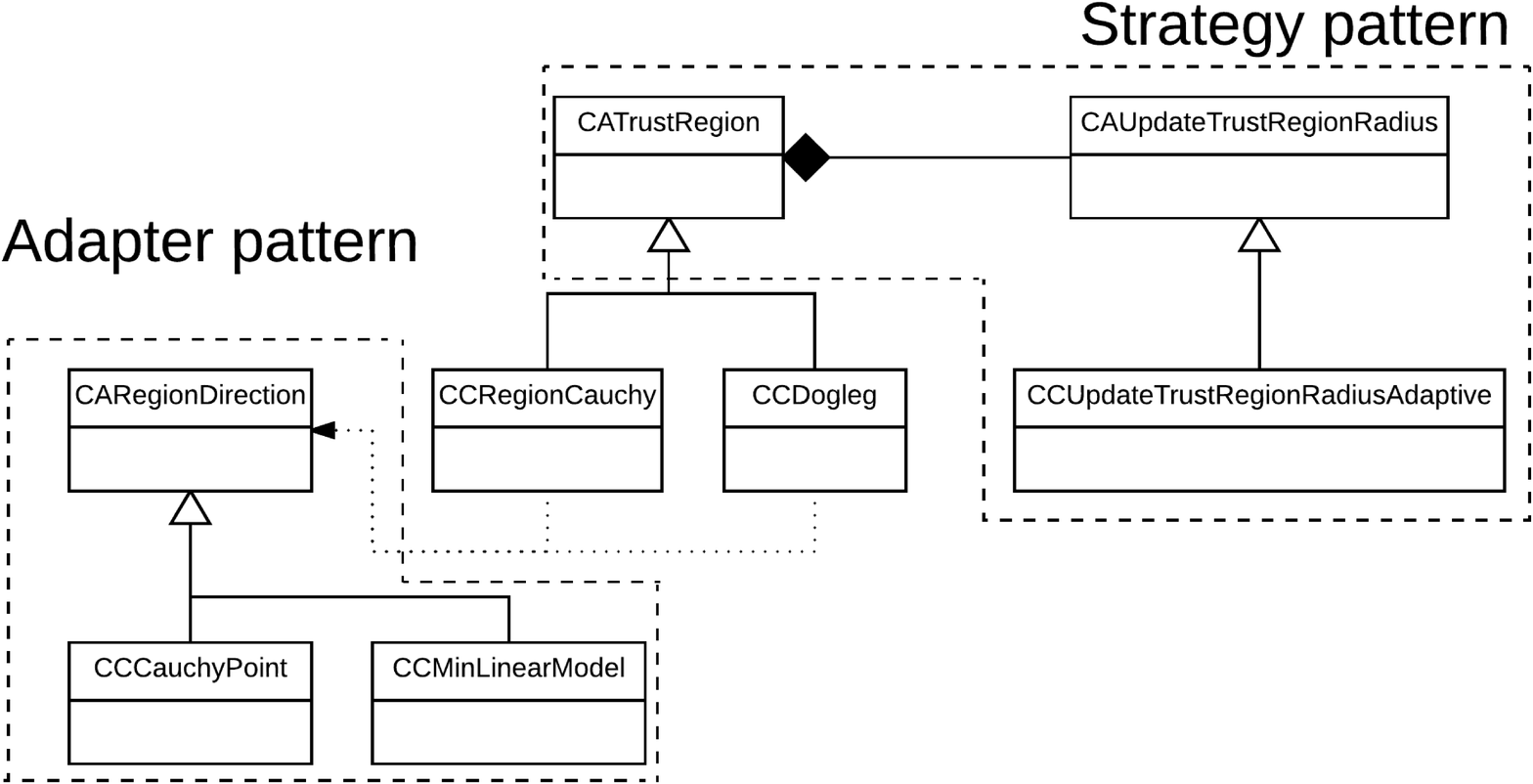}
  \caption{Software design for Trust Region Methods applying the
    Strategy and the Adapter patterns.}
  \label{Fig:trust_region_design}
\end{figure}

We apply the \textbf{Strategy} pattern \cite{Gamma:1995:BK} to
encapsulate the algorithms that solve equation
\eqref{eq:constrained_min_problem} in a family of classes that share
the same interface. The \textbf{Adapter} pattern helps us to reuse
algorithms in different contexts, in this case it is applied to reuse
the methods that solve nonlinear unconstrained optimisation problems
to update the descending direction in equation
\eqref{eq:constrained_min_problem}, see Figure
\ref{Fig:trust_region_design}.

The software design presented in this section belongs to the Newton
component: Newton direction from Figure
\ref{Fig:relations_concepts_algorithm}.

% \begin{figure}
%   \centering
%   \includegraphics[scale=0.38]{old_figures/bridgetemplatefacade}
%   \caption{Template Method, Facade and Bridge patterns.}
%   \label{Fig:bridge_template_facade}
% \end{figure}

\subsection{Objects Creation, the Abstract Factory and the Singleton
  Patterns}
\label{sub:objects_creation_the_abstract_factory_and_the_singleton_patterns}
\begin{figure}[H]
  \centering
  \includegraphics[width=0.95\textwidth]{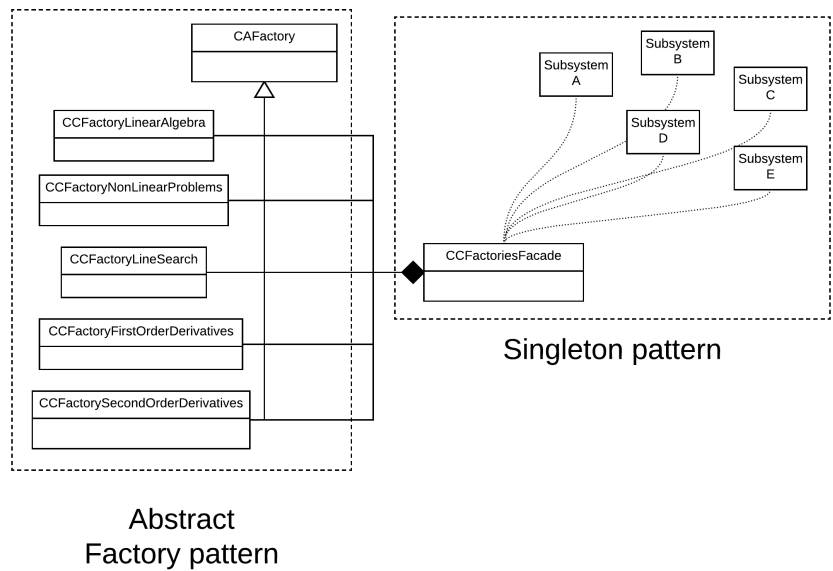}
  \caption{Software design implementing the Abstract Factory and the
    Singleton design patterns. Note that the prefix \texttt{CA} and
    \texttt{CC} are used to identify abstract and concrete classes,
    respectively.}
  \label{Fig:abstract_factory_and_singleton_patterns}
\end{figure}

In order to grant additional flexibility for the integration of new
routines or algorithms we apply the Abtract Factory pattern. This
pattern provides an interface to create families of objects without
specifying their concrete classes \cite{Gamma:1995:BK}. A single
global object implemented via the Singleton pattern
\cite{Gamma:1995:BK} is used as the interface for each of the
factories in the software design \cite{Serrato:2011:MTh}, see Figure
\ref{Fig:abstract_factory_and_singleton_patterns}.

\subsection{External Packages and the Adapter Pattern}
\label{sec:external_packages_and_the_adapter_pattern}
\begin{figure}[H]
  \centering
  \includegraphics[width=0.8\textwidth]{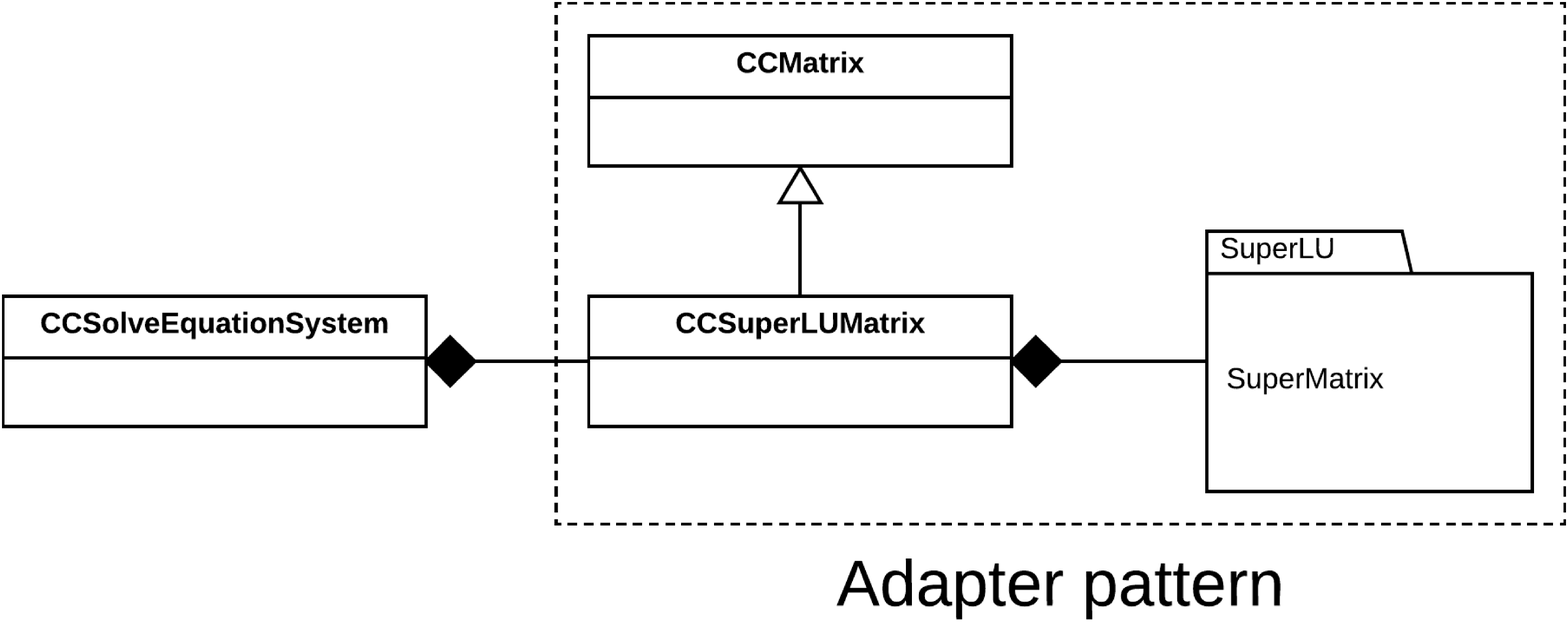}
  \caption{The adapter pattern allows to use external numerical
    libraries without modifying the existing source code.}
  \label{Fig:adapter_pattern}
\end{figure}

Newton-type methods solve a system of equations as part of its
computations, in order to include support for third-party
state-of-the-art numerical software libraries we use the Adapter
design pattern to convert an interface from an external package into
an interface expected by our software design, \cite{Gamma:1995:BK},
see Figure \ref{Fig:adapter_pattern}.

\subsection{Patterns summary}
\label{sec:patterns_summary}
In the previous section we identified and applied eight design
patterns from Gamma's book \cite{Gamma:1995:BK} for the development of
a software system design for Newton--type methods. In Table
\ref{Table:applied_design_patterns} we summarise the identified design
patterns and their implications in the designed software.

\begin{table}[H]
  \footnotesize
  %\centering
  \begin{tabular}{p{1.2cm} p{4.6 cm} p{4.6 cm}}
    \hline\noalign{\smallskip}
    \textbf{Desing Pattern} & \textbf{General intent} & \textbf{Specific use}\\
    \noalign{\smallskip}\hline\noalign{\smallskip}
    Bridge & Decouples abstraction from implementation such that both
             can vary independently & Facilitates the implementation
                                      of different versions of a
                                      method to target the needs of
                                      each user (debugging or
                                      production version) \\
    Facade & Provides a unified interface to a set of interfaces &
                                                                   Implements
                                                                   a
                                                                   simple
                                                                   and unique interface for the Newton component \\
    Adapter & Wraps an existing interface with a new one & Facilitates
                                                           the use of third-party
                                                           numerical
                                                           software
                                                           libraries. Also
                                                           used to
                                                           reuse Newton
                                                           direction
                                                           strategies
                                                           from Trust
                                                           Region Methods\\
    Abstract Factory & Provides an interface to create a family of
                       objects & Provides and interface to create and
                                 configure the objects of the software
                                 system. When combined with the
                                 Adapter pattern, they decouple the
                                 creation of external packages from
                                 the software design\\
    Singleton & Provides a global access points for an only-one
                instance of a class & Simplifies the the communication
                                      with the factory objects \\
    Template Method & Defines the skeleton of an algorithm and lets
                      sub-methods to implement each step & Defines the
                                                           general structure of the three presented Newton--type methods:
                                                           nonlinear,
                                                           trust
                                                           region and
                                                           line search\\
    State & Allows an object to modify its behaviour based on the
            internal state of the object & Switch the strategies to
                                           handle nonlinear functions
                                           and its derivatives based
                                           on the problem to solve\\
    Strategy & Defines a family of algorithms and encapsulate them to
               make them interchangeable & Encapsulates the step-length test condition and
                                           the computation of the step-length methods\\
    \hline\noalign{\smallskip}
  \end{tabular}
  \caption{Applied design patterns.}
  \label{Table:applied_design_patterns}
\end{table}

\section{Evaluation of the Software Design}
\label{sec:newton_type_methods_software_evaluation}
In order to evaluate the quality of our software design we applied
Martin's metrics \cite{Martin:2003:BK}. These metrics consider the
dependency between subsystems to compute the instability and
abstractness of the software design. These measurements provide
insightful information regarding the adaptation, extension and
reusability capacities of the software design.

We identified 17 sub-systems or packages in our software system. For
each of these packages we measured their instability and abstractness
as indicated by Martin in \cite{Martin:2003:BK}.

\subsection{Abstractness}
\label{subsec:abstraction}
The level of abstractness of a software package is given by the ratio
between the number of abstract classes in the package and the total
number of classes in the package, that is
\begin{equation}
  \label{eq:metric_abstractness}
  A = \frac{N_a}{N_c},
\end{equation}
where $N_a$ represents the number of abstract classes in the package
and $N_c$ represents the total number of classes in the package. When
$A=1$ then we have an abstract packages, conversely, when $A=0$ then
we have a concrete package.

The level of abstractness of a package is an indicator of its capacity
for extension and reuse.

\subsection{Instability}
\label{subsec:instability}
The instability of a package is given by the ratio
\begin{equation}
  \label{eq:metric_instability}
  I = \frac{C_e}{C_e + C_a},
\end{equation}
where $C_e$ is the number of classes inside the package that depend on
classes outside the package, and $C_a$ is the number of classes
outside the package that depend on classes inside the package
\cite{Martin:2003:BK}.

If $C_e=0$ then $I = 0$, therefore we have an stable package, on the
contrary, if $C_a=0$ then $I = 1$, which indicates that we have an
unstable package.

\subsection{The main sequence}
\label{subsec:the_main_sequence}
Considering Martin's metrics, a good software design is that when
$D=0$, where $D = | A + I - 1 |$. $D$ is an indicator on the facility
of a package to be adapted or extended. A package value of $D=1$
indicates that the package is difficult to adapt or modify. In Table
\ref{Table:Martin_metric_applied} we present the obtained values for
the main packages of the developed software design.

\begin{table}[h]
  \centering
  \footnotesize
  \begin{tabular}{cccc}
    \hline
    \textbf{Package name} & $A$ & $I$ & $D$\\
    \hline
    \texttt{TrustRegionMethods} & 0.29 & 0.80 & 0.09\\
    \texttt{BaseArchitecture} & 0.75 & 0.31 & 0.06\\
    \texttt{LineSearchMethods} & 0.20 & 0.70 & 0.10\\
    \texttt{NonlinearMethods} & 0.29 & 0.46 & 0.25\\
    \hline
  \end{tabular}
  \caption{Martin's metric applied to the main packages of the
    architecture.}
  \label{Table:Martin_metric_applied}
\end{table}

We observe that $D=0.06$ for the package \texttt{BaseArchitecture}, it
indicates that the package is mostly abstract and has low dependency
with other packages, therefore the package is easy to extend or
adapt. Note that the \texttt{BaseArchitecture} package is part of the
foundation of the software design. The other packages are also close
to $D=0$.

\begin{figure}[h]
  \centering
  \includegraphics[width=0.5\textwidth]{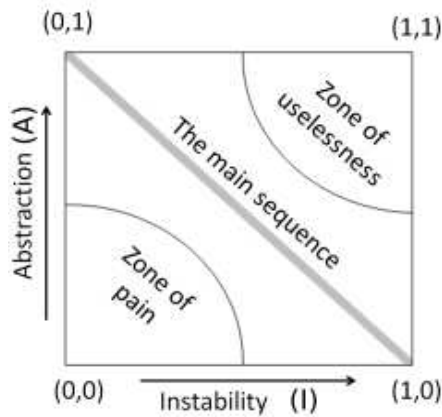}
  \caption{The main sequence. Note that $I$ and $A$ should add to one
    in order to stay close to the main sequence.}
  \label{Fig:the_main_sequence}
\end{figure}

The $D$ metric is better understood as a function of $I$ and $A$ as
shown in Figure \ref{Fig:the_main_sequence}. The closer the points
representing the packages are to the \textit{main sequence}, the more
easy to adapt or extend they are.

\section{Conclusions}
\label{sec:conclusions}
%% 12/junio/2016
%% Julio estoy actualizando (cambiando) las conclusiones
We presented a new methodology based on a domain analysis and a Scope,
Commonality and Variability analysis to transfer the knowledge of
scientific experts into simple, flexible and effective object-oriented
software designs. We applied the presented methodology to develop a
pattern-object-oriented software design for Newton--Types methods. We
evaluated the instability and the flexibility of the developed
software design by means of Martin's metric. The results shown that
the software design is stable enough to be extended without loss of
flexibility.

Newton's method has many applications in scientific computing; in this
work we used for solving optimisation problems. Another of its main
uses is the approximation of solutions of systems of equations arising
in simulation problems involving the finite difference method, the
finite element method or the finite volume method. In this regard, the
presented software design may be used as the core of an specialized
software for the simulation of physical phenomena or industrial
processes.

Regarding the identification of design patterns for the development of
scientific and engineering software, we successfully applied eight
patterns from Gamma's book \cite{Gamma:1995:BK}, namely: bridge,
facade, adapter, abstract factory, singleton, template method, state
and strategy. In this regard, a main contribution of this work is the
identification and application of the \textbf{state} pattern; after a
thorough search in the relevant literature we found no reports of the
application of this pattern for the development of scientific
software.

As part of our future work, in order to overcome the abstraction
penalty, introduced by the application of design patterns, in the code
performance, we are working in the use of parallel technologies, the
integration of third-party state-of-the-art software libraries and
code optimisation techniques for the development of high-performance
numerical software.

\section{Acknowledgements}
The first author wants to thank to CONACyT for partially funding the
development of this work as part of his master thesis project.

% ========================================================================
% Customising bibliography ------------------- BEGIN ---------------------
% ========================================================================
%\printbibliography[heading=bibintoc]
% Print bibliography from articles
%\printbibliography[heading=bibintoc, type=article,title={Articles}]
% Print bibliography from books
%\printbibliography[heading=bibintoc,type=book,title={Books}]
% Print bibliography from incollection
%\printbibliography[heading=bibintoc,type=incollection,title={In collection}]
% ========================================================================
% Customising bibliography ------------------- END -----------------------
% ========================================================================

% ========================================================================
% Customising bibliography [BibTex] ---------- BEGIN ---------------------
% ========================================================================
\bibliography{biblio}
\bibliographystyle{acm}
%\bibliographystyle{ieeetr}
%\bibliographystyle{plain}
%\bibliographystyle{alpha}
%\bibliographystyle{siam}
%\bibliography{biblio}
% ========================================================================
% Customising bibliography [BibTex] ---------- END -----------------------
% ========================================================================

%\end{multicols}

\end{document}